\documentclass{JAC2003}
\usepackage{amsmath}
\usepackage{amssymb}
\usepackage{graphicx}
\usepackage{booktabs}

\begin{document}

\title{INVESTIGATING THE ONE-PHOTON ANNIHILATION CHANNEL IN AN $e^-e^+$
PLASMA CREATED FROM VACUUM IN STRONG LASER FIELDS}
\author{D.B. Blaschke, University of Wroc{l}aw, 50-204 Wroc{l}aw, Poland;
JINR, 141980 Dubna, Russia \\
G.~R\"opke, Institut f\"ur Physik, Universit\"at Rostock, D-18051 Rostock, 
Germany\\
V.V. Dmitriev, S.A. Smolyansky\thanks{%
smol@sgu.ru}, A.V. Tarakanov, Saratov State University, 410026 Saratov,
Russia}
\maketitle

\begin{abstract}
It is well known that in the presence of strong external electromagnetic
fields many processes forbidden in standard QED become possible.
One example is the one-photon annihilation process considered recently
by the present authors in the framework of a kinetic approach to the
quasiparticle $e^{-}e^{+}\gamma$ plasma created from vacuum in the focal
spot of two counter-propagating laser beams.
In these works the domain of large values of the adiabaticity parameter
$\gamma\gg1$ (corresponding to multiphoton processes) was considered.
In the present work we estimate the intensity of the radiation stemming
from photon annihilation in the framework of the effective mass model
where $\gamma\lesssim 1$, corresponding to large electric fields
 $E\lesssim E_c=m^2/e$ and high laser field frequencies $\nu\lesssim m$
(the domain characteristic for X-ray lasers of the next generation).
Under such limiting conditions the resulting effect is sufficiently large
to be accessible to experimental observation.
\end{abstract}

\section{INTRODUCTION}

The planned experiments \cite{1} for the observation of an $e^{-}e^{+}$
plasma created from the vacuum in the focal spot of two counter-propagating
optical laser beams with the intensity $I\gtrsim 10^{21}$ W/cm$^{2}$ raises
the problem of an accurate theoretical description of the experimental
manifestations of the dynamical Schwinger effect \cite{2}, see also 
Refs.~\cite{3,4,5}. 
The existing prediction \cite{6} in the domain of strongly sub-critical fields 
$E\ll E_{c}=m^{2}/e$ of a considerable number of secondary annihilation photons
is not rather convincing because it is based on the S-matrix approach for the 
description of quasiparticle excitations in the presence of a strong external 
electric field. 
In particular, this approach does not take into account vacuum polarization 
effects. 
Apparently, an adequate approach for the description of vacuum excitations in 
strong electromagnetic fields is a kinetic theory in the quasiparticle
representation. 
The simplest kinetic equation (KE) of such type for the $e^{-}e^{+}$ subsystem 
has been obtained for the case of linearly polarized, time dependent and 
spatially homogeneous electric fields \cite{2}. 
Some generalizations of the KE in the fermion sector have been worked out in
Refs.~\cite{7,7a,7b}. 
It can be expected, that electromagnetic field fluctuations
of the $e^{-}e^{+}$ plasma are accompanied by the generation of real photons
which can be registered far from the focal spot. 
The first two equations of the BBGKY chain for the photon sector of the 
$e^{-}e^{+}$ plasma were obtained in \cite{8,8a}. 
This level is sufficient for the kinetic description of the one-photon 
annihilation. 
In the presence of an external field such process is not forbidden \cite{9}. 
In the works \cite{8,8a} it was shown that the spectrum of the secondary 
photons in the low frequency domain $k\ll m$ has the character of the flicker 
noise. 
In the work \cite{10} the inclusion of vacuum polarization effects in the 
one-photon radiation spectrum led to an essential change of the photon KE 
which was investigated in a broad spectral band including the annihilation 
domain $\nu \sim 2m$.
First we have considered the domain of large adiabaticity parameters 
$\gamma \gg 1$, where the photon radiation from the focal spot turns out to be 
very small.
However, the tendency of the effect to grow for $\gamma \rightarrow 1$ has
been discovered. 
This is just the domain of practical interest for parameters of modern lasers.

In the present work the effective mass model is considered which allows to
investigate the photon radiation in the domain of rather strong fields
not restricted to specific values of the adiabacity parameter.
Some crude estimations in the framework of this model \cite{9} lead to an
unexpectedly large total photon production intensity.

\section{EFFECTIVE MASS MODEL}

We will proceed from the photon kinetic equation
\begin{gather}
\dot{F}(\vec{k},t)=\frac{e^{2}}{4k(2\pi )^{3}}\int^{t}dt^{\prime }\int
d^{3}pe^{-i\theta (\vec{p},\vec{p}+\vec{k},\vec{k};t^{\prime },t)}\times
\notag \\
\times K(\vec{p},\vec{p}+\vec{k},\vec{k};t^{\prime },t)f(\vec{p},t^{\prime
})f(\vec{p}+\vec{k},t^{\prime })+c.c.
\label{1}
\end{gather}
for the one-photon annihilation mechanism taking into account
vacuum polarization effects in the low density approximation \cite{10}.
In Eq. (\ref{1}) $F(\vec{k},t)$ and $f(\vec{p},t)$ are the photon and electron
(positron) distribution functions, respectively, $\vec{k}$ is the wave
vector of the radiated photon and
\begin{equation}
\theta (\vec{p}_{1},\vec{p}_{2},\vec{k};t^{\prime
},t)=\int\limits_{t^{\prime }}^{t}d\tau \left[ \omega (\vec{p}_{1},\tau
)+\omega (\vec{p}_{2},\tau )-k\right]  \label{2}
\end{equation}%
is the high frequency phase.

The two-time convolution $K(\vec{p},\vec{p}+\vec{k},\vec{k};t^{\prime },t)$
of the four-spinors is a slowly varying function of its variables and can be
replaced by its average $K\rightarrow K_{0}\sim 1$, which is sufficient for 
coarse estimations.

The effective mass model \cite{9} is based on the approximation%
\begin{eqnarray}
\omega (\vec{p},t) &=&\sqrt{m^{2}+\left( \vec{p}-e\vec{A}(t)\right) ^{2}}%
\rightarrow  \notag \\
&\rightarrow &\omega _{\ast }(p)=\sqrt{m_{\ast }^{2}+\vec{p}^{2}},  \label{3}
\end{eqnarray}%
with the effective mass defined by the relation
\begin{eqnarray}
m_{\ast }^{2} &=&m^{2}+e^{2}\left\langle \vec{A}^{2}(t)\right\rangle
=m^{2}+e^{2}E_{0}^{2}/2\nu ^{2}=  \notag \\
&=&m^{2}(1+1/2\gamma ^{2}),  \label{4}
\end{eqnarray}%
where $\left\langle ...\right\rangle $ denotes the time averaging operation,
$\nu $ is the frequency of the periodic laser field and $E_{0}$
is its field strength amplitude, $\gamma =(E_{c}/E_{0})(\nu /m)$ is the
adiabaticity parameter.

In this approximation the phase (\ref{2}) becomes monochromatic
\begin{equation}
\theta (\vec{p}_{1},\vec{p}_{2},\vec{k};t^{\prime },t)=\Omega _{\ast }(\vec{p%
}_{1},\vec{p}_{2},\vec{k})(t-t^{\prime }),  \label{5}
\end{equation}
\begin{equation}
\Omega _{\ast }(\vec{p}_{1},\vec{p}_{2},\vec{k})=\omega _{\ast }(\vec{p}%
_{1})+\omega _{\ast }(\vec{p}_{2})-k,  \label{6}
\end{equation}%
i.e. the approximation (\ref{3}) leads to a suppression of multiphoton
processes (it corresponds to the large values of the adiabacity parameter $%
\gamma \gg 1$) and the mismatch (\ref{6}) can be compensated by the
harmonics of the fermion distribution functions in Eq. (\ref{1}) only.

The inspection of the fermion distribution function shows, in particular,
that it oscillates basically with twice the laser frequency
\begin{equation}
f(\vec{p},t)=\frac{1}{2}\bar{f}(\vec{p})\left[ 1-\cos (2\nu t)\right] .
\label{eq:7}
\end{equation}%
The substitution of Eqs. (\ref{5}) and (\ref{eq:7}) into the kinetic equation
(\ref{1}) allows to perform the time integration, leading to the appearance
of two harmonics in the radiation spectrum only (the 2$^{\rm nd}$ and the 
4$^{\rm th}$),
\begin{equation}
\dot{F}(\vec{k},t)=-A^{(2)}(\vec{k})\cos (2\nu t)+A^{(4)}(\vec{k})
\cos (4\nu t),
\label{eq:8}
\end{equation}
{\small
\begin{equation}
A^{(2)}(\vec{k})=
\frac{\pi ^{2}K_{0}\alpha }{2k}\int \frac{d^{3}\vec{p}}{(2\pi )^{3}}
\bar{f}(\vec{p})\bar{f}(\vec{p}+\vec{k})\delta \left( 2\nu -\Omega_{\ast }
\right)~,
\label{eq:9}
\end{equation}
\begin{equation}
A^{(4)}(\vec{k})=
\frac{\pi ^{2}K_{0}\alpha }{8k}\int \frac{d^{3}\vec{p}}{(2\pi )^{3}}
\bar{f}(\vec{p})\bar{f}(\vec{p}+\vec{k})\delta \left( 4\nu -\Omega_{\ast }
\right)~.
\label{10}
\end{equation}
}
It is important that a constant component is absent here, because the
mismatch (\ref{6}) could not be compensated in this case by other sources of 
the time dependence on the r.h.s. of Eq.~(\ref{1}).\footnote{This is in 
contrast to the case $\gamma \gg 1$, where accounting for multi-photon 
processes in the phase (\ref{2}) leads to a constant component \cite{10}.}

Thus, in the case of the infinite system the solution (\ref{eq:8}) can be
interpreted as "breathing" of the photon subsystem.
However, the situation is changed, when the generation of the 
$e^{-}e^{+}\gamma $
plasma is considered in a small spatial domain of the focal spot with volume
$\sim \lambda ^{3}$ due to the vacuum condition of the absence of the
$e^{-}e^{+}\gamma $ plasma in the initial moment of switching on the laser 
field.
In this case one can expect, that all annihilation photons generated in the 
first half-period of the field will leave the volume of the system and 
therefore in the next half-period the reverse process (photon transformation 
to $e^{-}e^{+}$ plasma) will be impossible.
Such a mechanism leads to a pulsation pattern for the photon radiation from the
focal spot.
It corresponds to introducing the condition of a positive definite photon 
production rate on the r.h.s of Eq. (\ref{eq:8}).

For estimations of the amplitudes (\ref{eq:9}), (\ref{10}) let us introduce the
additional model approximation in the spirit of the model (\ref{3})
\begin{equation}
\omega _{\ast }(\vec{p}+\vec{k})\rightarrow \omega _{\ast }(p,k)=\sqrt{%
\omega _{\ast }^{2}(p)+k^{2}}  \label{11}
\end{equation}%
and the isotropisation condition 
$\bar{f}(\vec{p}+\vec{k})\rightarrow \bar{f}(p+k)$.
The integrals on the r.h.s of Eqs. (\ref{eq:9}), (\ref{10}) can then be 
calculated.
For example,
\begin{gather}
A^{(2)}(\vec{k})=\frac{K_{0}\alpha }{4k}\bar{f}(p_{0})\bar{f}(p_{0}+k)
\frac{\omega _{\ast }(p_{0})\omega _{\ast }(p_{0},k)}{\omega _{\ast
}(p_{0})+\omega _{\ast }(p_{0},k)}p_{0},  \label{12}
\end{gather}%
where
\begin{equation}
p_{0}=\sqrt{\frac{4\nu ^{2}(k+\nu )^{2}}{(k+2\nu )^{2}}-m_{\ast }^{2}}
\label{13}
\end{equation}%
is the root of the equation $\Omega _{\ast }-2\nu =0$. From Eq. (\ref{13})
it is follows the threshold condition\footnote{
A similar effect was found first in the theory describing the absorption of 
a weak signal by the $e^{-}e^{+}$ plasma created from vacuum \cite{11}.}
\begin{equation}
\frac{2\nu (k+\nu )}{k+2\nu }\geqslant m_{\ast }~.
\label{14}
\end{equation}%
This condition is rather nontrivial because the effective mass (\ref{4}) 
depends also on $\nu $.
The minimal permissible value $\nu =2m_{\ast }$ corresponds to $k=0$.
For the 4$^{\rm th}$ harmonic the threshold value falls to $\nu =m_{\ast }$,
which is close to the parameters of the XFEL \cite{12}.

The $1/k$ - dependence on the r.h.s. of Eq. (\ref{12}) corresponds to the 
flicker noise. 
This feature in the spectrum of radiated annihilation photons has been
found first in Ref.~\cite{8}.

The number of photons with the frequency $k$ lying in the interval $[k,k+dk]$
and radiated from the focal spot with the volume $\lambda ^{3}=\nu ^{-3}$ per
time interval is
\begin{equation}
\frac{d^{2}N}{dtdk}=\frac{8\pi k^{2}}{\nu ^{3}}\dot{F}(\vec{k},t).
\label{15}
\end{equation}%
The fraction on the r.h.s. of Eq. (\ref{12}) is a slow function of the
frequencies $k$ and $\nu $ and for the sake of a preliminary estimation it 
can be replaced by $m_{\ast }/2$.
For the 2$^{\rm nd}$ harmonic we then obtain from Eqs. (\ref{12}) and 
(\ref{15})
\begin{equation}
\frac{d^{2}N^{(2)}}{dtdk}=
\frac{2\pi \alpha K_{0}km_{\ast}}{\nu^{3}}\bar{f}(p_{0})\bar{f}(p_{0}+k)p_{0}~.
\label{16}
\end{equation}

As a representative characteristics of the effectiveness of the radiation from
the focal spot domain we will consider the total photon number per time 
interval,
\begin{equation}
\dot{N}^{(2)}=\frac{2\pi \alpha K_{0}m_{\ast }}{\nu ^{3}}\int
\limits_{0}^{k_{\max }}dk~k~\bar{f}(p_{0})\bar{f}(p_{0}+k)p_{0}~.
\label{17}
\end{equation}
The electron and positron distribution functions entering here are defined as
the solutions of the corresponding non-perturbative kinetic equation \cite{2,7}
describing vacuum creation of $e^{-}e^{+}$ pairs under the action of a
strong, time dependent electric field of a standing wave of two
counter propagating laser beams.
The cutoff parameter $k_{\max }=2m_{\ast }$ is introduced in order to take into
account the annihilation photons in the radiation spectrum.

The fermion distribution function $f(\vec{p},t)$ is a rapidly decreasing
function with its maximum in the point $\vec{p}=0$ \cite{3}.
On this basis for a rough estimation one can put $p_{0}=0$ in the arguments of 
these functions on the r.h.s. of Eq. (\ref{17}),
\begin{equation}
\dot{N}^{(2)}=\frac{2\pi \alpha K_{0}m_{\ast }}{\nu ^{3}}\bar{f}(0)
\int\limits_{0}^{k_{\max }}dk~k~\bar{f}(k)p_{0}~,
\label{18}
\end{equation}%
where according to Eq. (\ref{13})
\begin{eqnarray}
p_{0}(k) &=&\frac{m_{\ast }^{2}}{k+4m_{\ast }}
\sqrt{48+56\frac{k}{m_{\ast }}+15\frac{k^2}{m^2_{\ast }}} \simeq  \notag \\
&\simeq &\frac{m}{4}\sqrt{48+56\frac{k}{m_{\ast }}},  \label{19}
\end{eqnarray}%
because the small $k_{\max }\ll m_{\ast }$ is effective in the integral
(\ref{18}).
As the result, we obtain the following order of magnitude estimate
\begin{equation}
\dot{N}^{(2)}\sim \alpha m_{\ast }\bar{f}^{2}(0)~.
\label{20}
\end{equation}

For the XFEL\ parameters $E_{0}=0.24E_{c}$ and $\lambda =15$ nm \cite{12} we
have according to the kinetic theory in the $e^{-}e^{+}$ sector
$\bar{f}(0)\sim 10^{-2}$.
From Eq.~(\ref{20}) then follows
\begin{equation}
\dot{N}^{(2)}\sim 10^{17}~~\mathrm{s}^{-1}~.
\label{21}
\end{equation}
For the 4$^{\rm th}$ harmonic with the oscillation amplitude (\ref{10}) the
threshold for the generation of annihilation photons is lowered (see discussion
after Eq.~(\ref{14})) but the intensity of the photon radiation is also lowered
so that the order of magnitude of (\ref{21}) remains unchanged.

\section{SUMMARY}

We have considered the effective mass model \cite{9} which allows a rather
simple solution of the kinetic equation describing (in the framework of the
one-photon annihilation mechanism) the photon radiation from the focal spot
of two counter-propagating laser beams.
This simple model leads to considerable depletion of the spectrum of parametric
oscillations of the $e^{-}e^{+}\gamma $ plasma: 
only the 2$^{\rm nd}$ and 4$^{\rm th}$ harmonics remain due to the condition 
of the absence of a constant component in the photon production rate.
Thus a compensation of the mismatch (\ref{6}) is possible by means of these
two harmonics only.
The domain of applicability of this model is limited to the X-ray domain of
the laser radiation.
The model suggests a high integral luminosity of $\sim 10^{15}$ photons per sec
from the focal spot.
Other features of the model are: the $1/k$-behavior in the infrared
domain $k\ll m$ (the flicker noise of electrodynamic nature) and the threshold
effect.
These results are encouraging for a further detailed study of the photon 
kinetics on the basis of the one-photon annihilation mechanism in the domain 
of small adiabacity parameters $\gamma \lesssim 1$.

\section{ACKNOWLEDGEMENTS}

We thank our colleagues G.~Gregori, C.D.~Murphy, A.V.~Prozorkevich, 
C.D.~Roberts and S.~Schmidt for their collaboration. 
A.M.~Fedotov, D.~Habs, B.~K\"ampfer, H.~Ruhl and R.~Sauerbrey are 
acknowledged for their continued interest in our work and valuable discussions.

\end{document}